\documentclass{emulateapj}

\usepackage[usenames]{color}

\shorttitle{First {\it Kepler} results on RR\,Lyrae stars}
\shortauthors{Kolenberg et al.}

\begin{document}

\title{First {\it Kepler} results on RR\,Lyrae stars}

\author{K. Kolenberg\altaffilmark{1}, R. Szab\'o\altaffilmark{2}, D.~W. Kurtz\altaffilmark{3}, R.~L. Gilliland\altaffilmark{4}, J. Christensen-Dalsgaard\altaffilmark{5}, H. Kjeldsen\altaffilmark{5}, T.~M. Brown\altaffilmark{6}, J.~M. Benk\H{o}\altaffilmark{2}, M. Chadid\altaffilmark{7}, A. Derekas\altaffilmark{2,8}, M. Di Criscienzo\altaffilmark{9}, E. Guggenberger\altaffilmark{1}, K. Kinemuchi\altaffilmark{10,11}, A. Kunder\altaffilmark{12}, Z. Koll\'ath\altaffilmark{2}, G. Kopacki\altaffilmark{13}, P. Moskalik\altaffilmark{14}, J.~M. Nemec\altaffilmark{15}, J. Nuspl\altaffilmark{2}, R. Silvotti\altaffilmark{16}, M.D. Suran\altaffilmark{17}, W.~J. Borucki\altaffilmark{18}, D. Koch\altaffilmark{18}, J.~M. Jenkins\altaffilmark{19}}

\altaffiltext{1}{Institut f\"ur Astronomie, Universit\"at Wien, T\"urkenschanzstrasse 17, A-1180 Vienna, Austria} 
\altaffiltext{2}{Konkoly Observatory, H-1525 Budapest, P.O. Box 67, Hungary}
\altaffiltext{3}{Jeremiah Horrocks Institute of Astrophysics, University of 
Central Lancashire, Preston PR1\,2HE, UK}
\altaffiltext{4}{Space Telescope Science Institute, 3700 San Martin Drive, Baltimore, MD 21218, USA}
\altaffiltext{5}{Department of Physics and Astronomy, Aarhus University, DK-8000 Aarhus C, Denmark}
\altaffiltext{6}{Las Cumbres Observatory Global Telescope, Goleta, CA 
93117, USA}
\altaffiltext{7}{Observatoire de la C\^ote d'Azur, Universit\'e Nice Sophia-Antipolis, UMR 6525, Parc Valrose, 06108 Nice Cedex 02, France}
\altaffiltext{8}{Sydney Institute for Astronomy, School of Physics A28, University of Sydney, NSW 2006, Australia}
\altaffiltext{9}{INAF-Osservatorio Astronomico di Roma, Via Frascati 33,00040 Monte Porzio Catone, Rome, Italy}
\altaffiltext{10}{Department of Astronomy, University of Florida, 211 Bryant Space Science Center, Gainesville, Florida 32611, USA}
\altaffiltext{11}{Departamento de Astronom\'{i}a, Universidad de Concepci\'on, Casilla 160-C,
Concepcion, Chile}
\altaffiltext{12}{Cerro Tololo Inter-American Observatory, National Optical Astronomy Observatory, La Serena, Chile}
 \altaffiltext{13}{Instytut Astronomiczny Uniwersytetu Wroc\l{}awskiego, Kopernika 11,
       51-622 Wroc\l{}aw, Poland}
\altaffiltext{14}{Copernicus Astronomical Centre, ul. Bartycka 18, 00-716 Warsaw, Poland}
\altaffiltext{15}{Department of Physics \& Astronomy, Camosun College, Victoria, British Columbia, Canada}
\altaffiltext{16}{INAF-Osservatorio Astronomico di Torino, Strada dell'Osservatorio 20, 10025 Pino Torinese, Italy}
\altaffiltext{17}{Astronomical Institute of the Romanian Academy, Str. Cutitul de Argint, 5, RO 40557, Bucharest, Romania}
\altaffiltext{18}{NASA Ames Research Center, MS 244-30, Moffett Field, CA 94035, USA}
\altaffiltext{19}{SETI Institute/NASA Ames Research Center, MS 244-30, Moffett Field, CA 94035, USA}

\begin{abstract}

We present the first results of our analyses of selected RR\,Lyrae stars
for which data have been obtained by the {\it Kepler Mission}. As expected, we find a 
significant fraction of the RRab stars to show the Blazhko effect, a still 
unexplained phenomenon that manifests itself as periodic amplitude and phase 
modulations of the light curve, on time scales of typically tens to hundreds of days. The long 
time span of the {\it Kepler Mission} of 3.5\,yrs, and the unprecedentedly high precision of 
its data provide a unique opportunity for the study of RR\,Lyrae stars. Using data 
of a modulated star observed in the first roll as a showcase, we discuss the data, 
our analyses, findings, and their implications for our understanding of RR\,Lyrae 
stars and the Blazhko effect.
With at least 40\% of the RR~Lyrae stars in our sample showing modulation, we confirm the high incidence rate that was only found in recent high-precision studies.  Moreover, we report the occurrence of additional frequencies, beyond the main pulsation mode and its modulation components.  
Their half-integer ratio to the main frequency is reminiscent of a
 period doubling effect caused by resonances, observed for the first time in RR\,Lyrae stars.

\end{abstract}


\section{Introduction}

RR\,Lyrae stars are low-mass stars that have evolved away from the main sequence and are 
burning Helium in their core. Their evolutionary stage makes them useful tracers 
of galactic evolution. Like the Cepheids, they obey a period-luminosity-color 
relation and are used as distance indicators. RR\,Lyrae stars have typical periods 
of $\sim$0.2 to $\sim$1\,d, amplitudes in the optical of 0.3 up to 2\,mag, and spectral types 
of A2 to F6. Most RR\,Lyrae stars pulsate in the radial fundamental mode (RRab 
stars), the radial first overtone (RRc stars) and, in some cases, in both modes 
simultaneously (RRd stars). A few RR~Lyrae stars are suspected to be pulsating in higher-order radial overtone modes \citep{OlM09}.

\subsection{The Blazhko effect}

A large fraction of the RR\,Lyrae stars show a nearly periodic modulation of their 
light curve amplitudes and phases on time scales of typically tens to hundreds of 
days. This so-called Blazhko effect \citep{Bla07} is one of the most stubborn unsolved 
problems of the theory of radial stellar pulsations. Since it was discovered more 
than a century ago, many hypotheses have been proposed to explain the modulation. 
Over the past decade, focus has predominantly been on resonance models and 
magnetic models. The oft-quoted resonance models \citep{VH98,DzM04} involve 
nonradial modes of low spherical degree (most likely $\ell=1$), whereas the 
magnetic model \citep {Shi00} proposes that a strong dipole magnetic field (of the 
order of 1 kG) inclined to the rotation axis deforms the main radial mode to have 
an additional nonradial quadrupole ($\ell$=2) component aligned with the magnetic 
field, thus generating the amplitude and phase modulation with the rotation period.

Recently, \citet{Cha04} and \citet{KB09} disproved the presence of a strong 
magnetic field with dipole-like geometry in RR\,Lyr, the prototype of the class, 
and for a sample of RR\,Lyrae stars respectively. This implies that the magnetic 
models in their present form need adaptation, or are not correct. 
\citet{Sto06} proposed a scenario that does not involve nonradial modes, but convective/magnetic interaction as a cause for modulation. The geometry of the magnetic field might be too complex to be detected. Importantly, \citet{JS09} showed that modulated stars change their mean global physical parameters (mean radius, luminosity and surface effective temperature) over the Blazhko cycle, and phase modulation of the pulsation is interpreted to be a consequence of period changes.

Thus far, none of the proposed models successfully accounts for all of the features observed in modulated RR\,Lyrae stars. 
The inspiration for new types of models should come from the observational side. 
The {\it Kepler mission} will be a prime driver for this.

\begin{figure*}
\label{fig1}
\epsscale{1.2}
\plotone{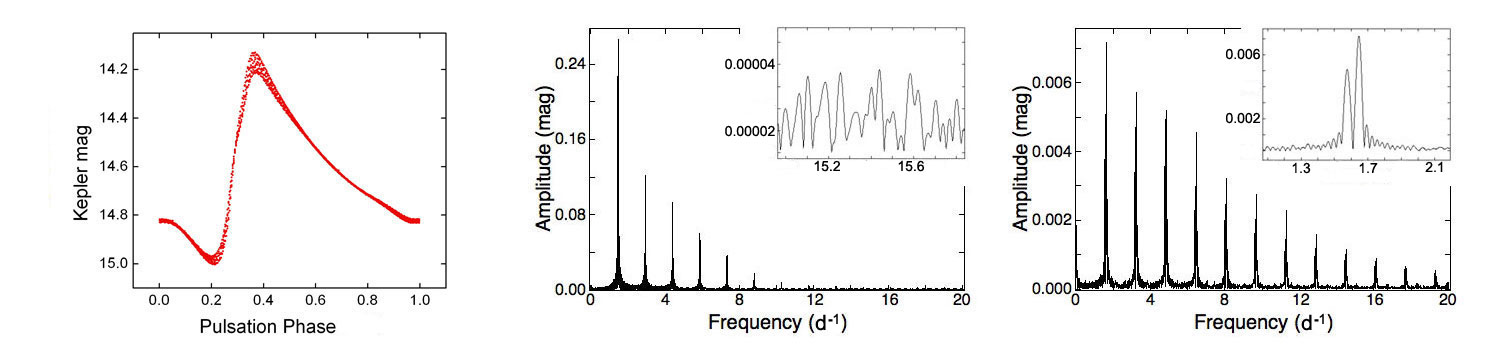}
\caption{{\bf Left:} Light curve of KIC\,5559631 folded with the main period of
0.620699\,d. The spread in the width of the light curve is not noise (which is much smaller
than the size of the data points), but shows the amplitude and phase modulation over the
27.6-d Blazhko cycle. {\bf Middle:} Fourier Transform of the raw data converted to the magnitude scale. The insert shows the noise level. {\bf Right:} Fourier Transform after prewhitening with the main frequency and its harmonics. The insert is a zoom, on the same scale, around the position of the main frequency and its harmonics.  Triplet components are clearly detected.}
\end{figure*}

\subsection{Targets}

To select the RR\,Lyrae targets for the {\it Kepler Mission}, we searched for all known 
RR\,Lyrae stars or possible candidates in the {\it Kepler} field. The list was compiled using all available variability information contained in the updated catalogs:
 GCVS \citep{gcvs}, ASAS North \citep{asas08}, ROTSE \citep{akerlof00}, and HAT \citep{hat04}. KIC-10 (KIC = {\it {\it Kepler} Input Catalog}) was used to get {\it Kepler} IDs, coordinates and 
magnitudes. The period range $0.15-1.0$\,d was searched in these catalogs. 
Light curve shapes and the log~P vs. $J-H$ diagram \citep{pm04} were 
also utilized for further selection. Finally, stars with 
close (bright) companions were excluded. RR\,Lyr itself, the prototype of the 
class, is also located in the {\it Kepler} field. 
Despite coordinated ground-based efforts we were not able to identify any Blazhko 
star, besides RR\,Lyr itself, in the {\it Kepler} field prior to the launch of the mission.

The {\it Kepler} magnitude $Kp$ (wide passband between $4300-9000$\,\AA) 
of the proposed targets lies in the range $7.9 \le Kp \le 17.4$.
A total of 57
RR\,Lyrae candidates, including some with uncertain classification, were proposed for observation. 
Of these, 48 stars 
were proposed in long cadence (one measurement every 30 minutes) for 90\,d and nine 
stars in short cadence (one measurement every minute) for 30\,d. A large fraction 
(23 out of 51) of the stars for which data were released at the time of writing 
turned out not to be RR\,Lyrae stars, but, e.g., eclipsing binaries and ellipsoidal variables. 

About 60\% of the survey targets will be proposed for continued observations after 
the survey period.

\section{Observations}

The asteroseismic data released to the KASC ({\it Kepler} Asteroseismic Science Consortium) at the time of this writing are long cadence observations. For stars with pulsation cycles of typically half a day this sampling is sufficient to 
obtain a good coverage of the light curve, but rapid changes and ``glitches" in 
the light curve -- as have been reported to occur in some RR\,Lyrae stars -- are 
missed by this sampling. All of the RR\,Lyrae stars with {\it Kepler} data have so far 
turned out to be pulsating predominantly in the radial fundamental mode (RRab).

Note that at the time of writing the pipeline for reducing the {\it Kepler} data is 
still being fine-tuned and tested. As a consequence, some low-frequency, 
low amplitude variability may be attributable to instrumental effects that will be 
overcome further into the mission. The results presented in this letter are valid 
within this minor constraint.

\section{An example: a Blazhko star observed with {\it Kepler}}

To illustrate the potential of {\it Kepler} photometry for RR\,Lyrae stars, we present the 
first frequency analysis results for the Blazhko star KIC\,5559631 (V783\,Cyg, see Fig.\,1). Before the {\it Kepler} observations the star was known to be an
RRab star \citep{gcvs} with a mean {\it Kepler} magnitude of
$m_{\rm K} = 14.643$ and a period $P_0=0.620699$\,d,
corresponding to a frequency $f_0=1.611087$\,d$^{-1}$. 
Blazhko modulation of the light curve was previously 
unknown. This star was observed in long cadence during the first roll of the 
{\it Kepler} survey phase between HJD\,2454964.0109 (2009 May 12) and HJD\,2454997.4812 
(2009 June 14). (The spacecraft "rolls" every three months to allow for continuous illumination of Kepler's solar arrays. The first such roll, ending the first quarter, occurred on June 18, 2009.)
Over the 33.5-d continuous run 1628 data points were obtained for 
the star. As a consequence of the sampling rate, the Nyquist frequency lies at 
24.5\,d$^{-1}$. 

A small linear trend was removed from the raw data and they were converted to the 
magnitude scale. We used several frequency analysis software packages, including 
Period04 \citep{LB05}, SigSpec \citep{ree07}, TiFrAn \citep{csu02}, CLEAN \citep{Rob87}, and PDM \citep{Ste78}. 

\begin{figure*} 
\label{fig2}
\epsscale{1.2}
\plotone{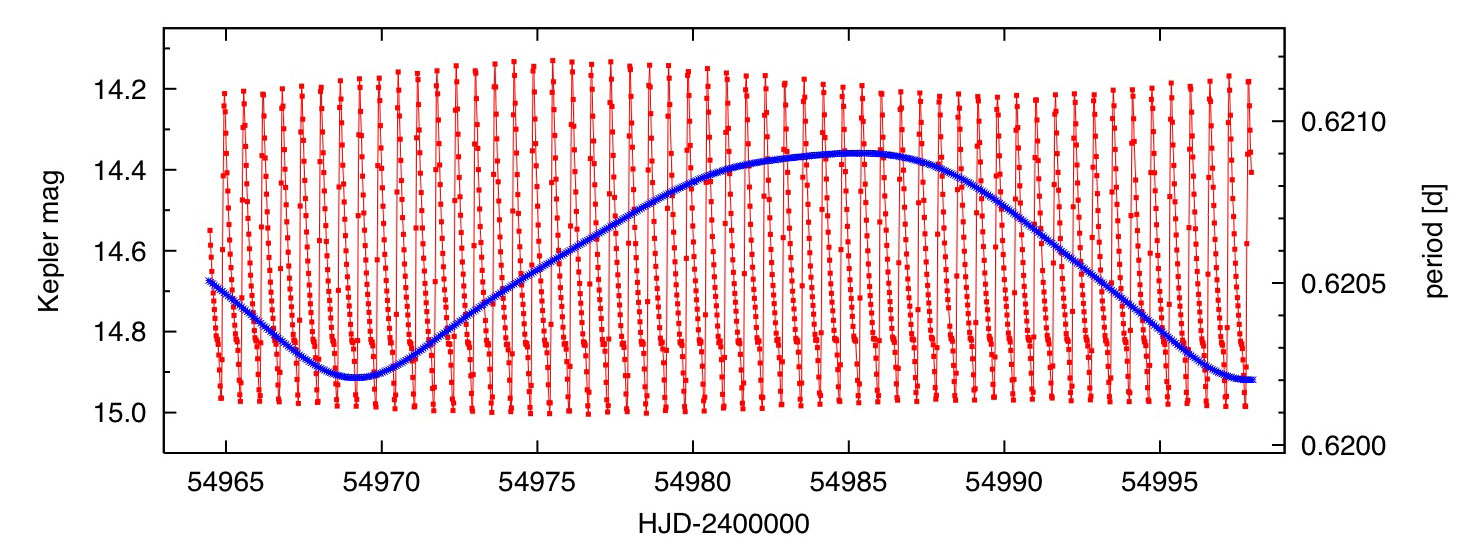}
\caption{Relationship between the period change and amplitude modulation over the Blazhko cycle for KIC\,5559631. The red dots show the Kepler data, the red line our multi-frequency fit to the data. The blue line shows an instantaneous period determined using the analytical signal method (units to the right side of the panel). The period change cannot be a Doppler shift caused by a companion, as that would imply an orbital velocity of the RR\,Lyrae star about the barycentre of about 200\,km\,s$^{-1}$, requiring an impossibly large mass for the companion.}
\end{figure*}

After prewhitening with the main frequency $f_0=1.61117 \pm 0.00003$ d$^{-1}$ and its significant harmonics (up to 25$f_0$; although this is past the Nyquist frequency, it is necessary, as the higher
harmonics are detected by their ``reflection" into the range below the Nyquist frequency by
beating with the sampling frequency), we clearly see the triplet frequencies that are a typical feature of 
Blazhko RR\,Lyrae stars (see Fig.\,1, right panel). The first side peak appears at $f_0+f_B=1.64735$ d$^{-1}$. 
This yields an initial value for the Blazhko frequency of $f_B=0.0362 \pm 0.0006$ d$^{-1}$ or a 
Blazhko period $P_B=27.6 \pm 0.5$ d.
Subsequent prewhitening reveals additional triplet frequencies equidistantly spaced around the main frequency and its harmonics. The triplet components are detected successively and significant up to the 23$^{\rm rd}$ order.
The Blazhko frequency $f_B$  itself is also detected directly from our data of  KIC\,5559631, though the peak in the Fourier spectrum at its location is very wide due to the short time base of the data.
Its detection implies that there is a variation of the mean brightness of the star with the Blazhko cycle.  
Though modulated RR\,Lyrae stars are commonly fitted with equidistant multiplet structures, it is worthwhile to test whether there are departures from equidistance. This will be done in our future analyses with a longer time base and hence better frequency resolution. 

We fitted the light curve of  KIC\,5559631 with the main frequency, its harmonics, the significant equidistant triplets, and the Blazhko frequency. The presence of further significant frequencies will be determined on the basis of future data of this target.

Fig.\,2 shows the light curve modulation over the Blazhko cycle for  KIC\,5559631.  An instantaneous period was determined continuously by the analytical signal method \citep{Koll02}, shown by the blue line in Fig.\,2.  The period change is non-sinusoidal and there is a considerable phase lag between the period and amplitude modulation. Similar behaviour was found by \citet{sza09} in the analysis of the CoRoT RR\,Lyrae Blazhko stars. 
This frequency variability will be fully characterized by the long-term {\it Kepler} observations
and will provide an additional new constraint on seismic models of RR\,Lyrae stars, since it
must be the result of global changes in the star over the Blazhko cycle.

\section{Results for the other RR\,Lyrae stars}

For the 28 RRab stars that have been observed by {\it Kepler} so far, we find periods in the range $0.43-0.68$\,d (main frequencies $f_0$ between 1.46 and 2.29\,d$^{-1}$), with observed amplitudes of the first Fourier component between 0.18 and 0.44 mag.  

\begin{figure*}
\epsscale{1.2}
\plotone{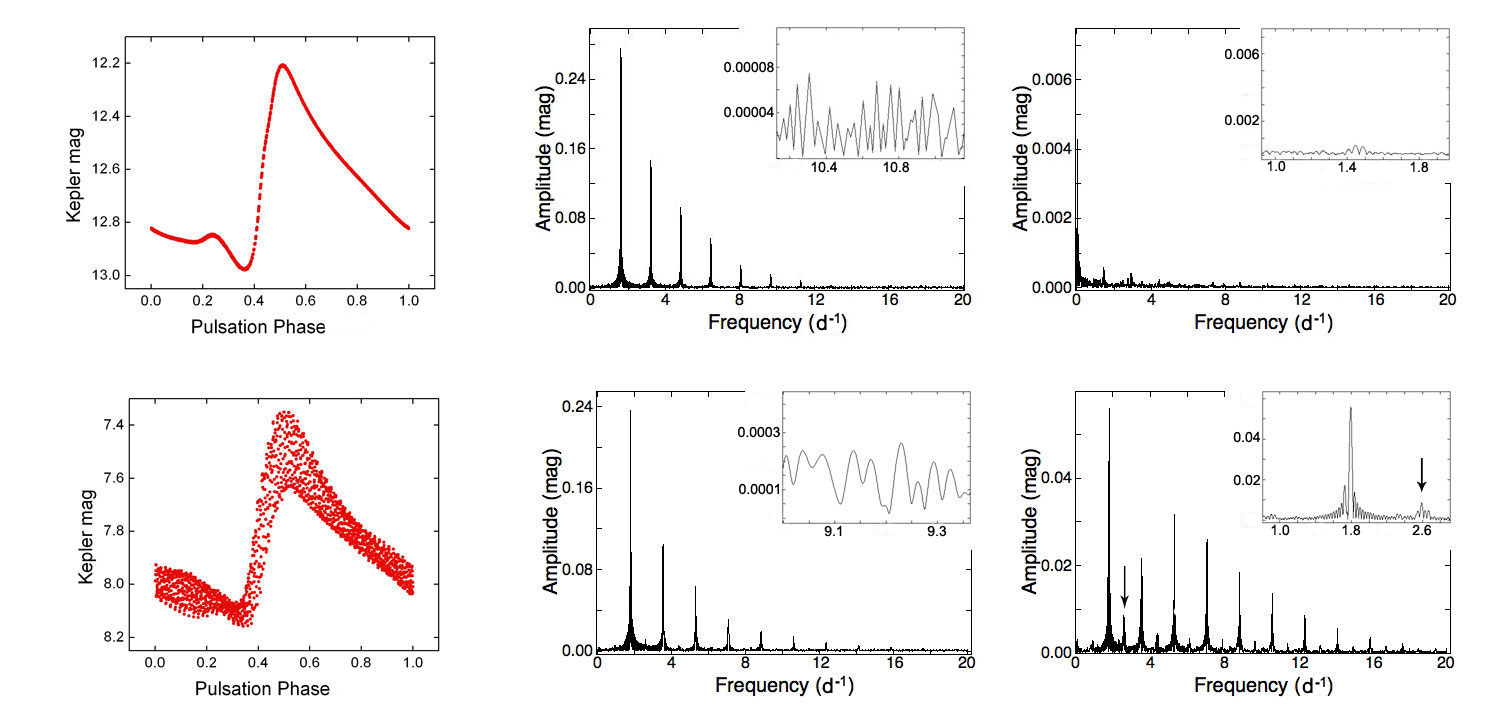}
\caption{{\bf Left:} Light curves of two {\it Kepler} RR\,Lyrae stars (KIC\,3733346 and KIC\,7198959) folded with the main period. {\bf Middle:} Fourier Transform of the raw data converted to the magnitude scale. The insert shows the noise level. {\bf Right:} Fourier Transform after prewhitening with the main frequency and its harmonics. The insert, on the same scale, 
is a zoom around the position of the main frequency and its harmonics.  For the modulated star (KIC\,7198959) multiplet components are clearly detected.  Arrows point at the highest peak connected to an additional frequency.
\label{fig3}}
\end{figure*}

Fig.\,3 shows example light curves of non-modulated (KIC\,3733346, NR\,Lyr) and
modulated (KIC\,7198959, RR\,Lyr itself) RR\,Lyrae stars in our sample and the matching Fourier transforms before and after subtraction of the main frequency and its significant harmonics. Note the occurrence of modulation peaks and additional frequencies in the Fourier spectrum of KIC\, 7198959. 

Some modulated stars in our sample already clearly show higher-order modulation side lobes 
in their frequency spectra. Their visibility can be attributed to the degree of amplitude/phase modulation (Szeidl \& Jurcsik 2009, Benk\H o et al. 2009) that we will quantify in our studies based on larger data sets.

Besides the modulation components that occur in multiplet structures around the main frequency and its harmonics, we also see additional frequencies that are not at the expected positions of radial modes in RR\,Lyrae stars (see also Section\,5.2).  They have significant amplitudes surpassing the mmag level (e.g., 8 mmag in KIC\,7198959). Their nature will be investigated on the basis of forthcoming data.

\section{Discussion}

\subsection{Modulation statistics}
An important statistic that should help constrain models for the Blazhko effect is 
its occurrence rate. 
In different stellar populations the relative number of Blazhko stars may vary \citep[see][and references therein]{Kov09}.
Previous estimates were that at least 20-30\% of the galactic RRab stars 
and 5-40\% of the RRc stars (the high RRc incidence rate was found in Omega Cen) are modulated \citep{Sze88,MP03,Miz03,MOl08}. Recent results from the 
Konkoly Blazhko Survey \citep{jur09} indicate a $\sim$47\% occurrence rate of the 
Blazhko effect (14 out of 30 RRab stars). From the excellent CoRoT data an even larger fraction of the 
observed RR\,Lyrae stars are modulated, though the sample size is very small 
\citep{sza09}. \citet{jur09} mention that the significant increase of the 
incidence rate is a consequence of the discovery of small-amplitude modulation 
\citep[see, e.g.,][]{jur06} that would not have been detected in any previous 
survey. 

In our sample of {\it Kepler} 
RR\,Lyrae targets, 11 out of 28 RRab stars are clearly modulated (with large amplitude) and 
for a few more we suspect small amplitude modulation. 
Though at this point our numbers are merely indicative, we obtain an incidence rate of at least 40\%, in agreement with the findings from recent high-precision studies \citep{jur09}, and higher than previously thought. 
Is every RRab star modulated?  What fraction of RR Lyrae stars are modulated to within certain detection limits? With unprecedented precision and the longer {\it Kepler} datasets we will investigate these questions, too.

\begin{figure}
\epsscale{1.2}
\plotone{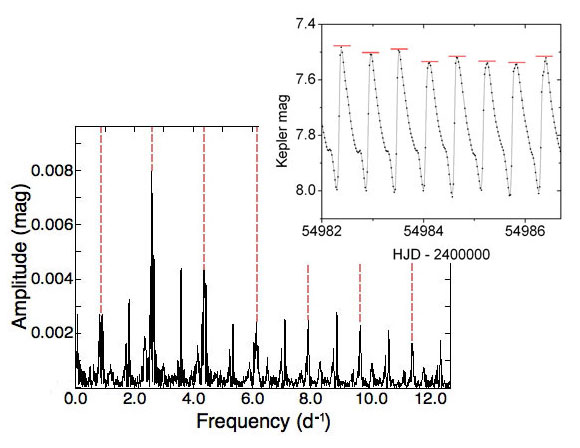}
\caption{Residual spectrum of KIC\,7198959 after subtraction of the main frequency $f_0$, its harmonics, and the triplet components.  The dashed lines mark the positions of the half-integer frequency components. The highest peak occurs at $\sim$3/2$f_0$. The frequency peaks in between the dashed lines are mostly higher-order multiplet components present in the spectrum around $f_0$ and its harmonics.  We did not subtract these yet, as the Blazhko period of KIC\,7198959 is only partially covered by these data. The insert shows the alternating heights of the maxima around minimum light for the star, connected to the additional frequencies.
\label{fig4}}
\end{figure}

\subsection{Additional frequencies}

For a few of our targets, we detect additional frequencies in our {\it Kepler} data that are not at the
expected positions of the radial modes in RR\,Lyrae stars.  However, they appear at frequency ratios close to 3/2, 1/2, 5/2, 7/2, with the radial fundamental mode $f_0$  (Fig.\,4). These ratios are reminiscent of resonance phenomena, specifically period doubling bifurcation as described by \citet{MB90} and \citet{BM92} for Cepheid and BL\,Her models \citep[see also Fig.\,3 in][for comparison with our Fig.\,4]{MB91}. Such period doubling would also result in alternating heights of maxima, as we observe in these targets (Fig.\,4).

In his recent review of the Blazhko effect, and our (lack of) understanding of the 
phenomenon, \citet{Kov09} stresses that more attention should again be paid to 
models based on {\it radial} mode resonances \citep[see, e.g.,][]{gor09}. After the models by \citet{Bor80} and 
\citet{Mos86} were dismissed by lack of confirmation from hydrodynamical models, 
most attention has been on resonances involving nonradial modes. 

In several recent studies of RR\,Lyrae stars additional frequencies were detected, e.g., in photometry of the RRd star AQ\,Leo \citep{Gru07} from the MOST satellite, and in the V1127\,Aql data from the CoRoT mission \citep{Cha09}. These frequencies were not found at positions where radial modes or radial resonances would occur. They may, therefore, be connected to nonradial modes.
Nonradial modes play a crucial role in several models for explaining the Blazhko modulation \citep[e.g.,][]{DzM04}. Their presence may, however, be unrelated to the Blazhko effect.  Despite several claims \citep{Ol99,Cha99} of detection and attempts \citep{Kol02} at their identification, there has been {\it no} unambiguous proof of nonradial modes in RR\,Lyrae stars so far.
We hope that 
through the ultra-precise data from the {\it Kepler mission} we will be able to obtain 
certainty about their existence.
Confirmation and 
exploration of nonradial pulsation would be extremely important to shift the 
current radial (1-dimensional) theoretical modeling to a nonradial (multi-dimensional) 
hydrodynamical description of RR\,Lyrae pulsations.


\subsection{Quality and advantages of {\it Kepler} photometry}

From the first releases of {\it Kepler} data it is obvious that {\it Kepler} photometry is 
stable enough to be tied to ground-based photometry, even if data are not taken 
simultaneously. Ground-based follow-up photometry and spectroscopy of selected 
{\it Kepler} targets is underway to improve the information in the {\it Kepler} Input 
Catalogue. High-resolution multi-color single-epoch imaging of the targets is 
important to mitigate the 4 arcsec sampling of the {\it Kepler} photometer. 
The latter may cause some problems concerning proper identification, determination 
of the total pulsational amplitude, and blending.

Although the CoRoT satellite is carrying out a similar mission to {\it Kepler}, and MOST 
is also delivering valuable results in asteroseismology, {\it Kepler} has significant  
advantages for the study of RR\,Lyrae stars. These include the large aperture of the 
primary mirror: 95-cm diameter (vs. 28-cm for CoRoT and 15-cm for MOST)  
allowing the observation of fainter stars and higher precision photometry, 
the potential for obtaining observations 
over a longer time base: 3.5-5\,yr (versus a maximum of 150\,d for CoRoT and about 
8 weeks for MOST in order to cover also longer cycles for 
modulated RR\,Lyrae stars (e.g., RR\,Lyr's 4-yr cycle), and the larger field of view: 
105 square degrees which allows the observation of a significantly large sample of 
RR\,Lyrae stars simultaneously. Finally, thanks to {\it Kepler}'s Earth-trailing orbit 
(vs. CoRoT's and MOST's low-Earth orbits), {\it Kepler} is spared from transits through 
the South Atlantic Anomaly (SAA).

\vspace{6mm}

\section{Summary}

The first releases of {\it Kepler} data have yielded light curves with unprecedented quality. 
In this Letter we have attempted to illustrate 
the tremendous potential of {\it Kepler} for studies of RR\,Lyrae stars.

The major results and prospects for further analysis can be summarized as follows:
\begin{itemize}
\item The fraction of modulated stars in our sample of 28 observed stars so far 
confirms that the Blazhko effect is much more common than previously thought \citep[see also][]{jur09}. We can already fix a {\it lower limit} of $\sim$40\% for the incidence rate in our sample.
The {\it Kepler} data will 
help to determine a lower limit for the modulation amplitude and to constrain the models.
\item We observe additional frequencies, beyond the main frequency and classically expected modulation components, in the frequency spectra of a number of modulated RR\, Lyrae stars.  Their ratio to the main frequency suggests resonance effects, in particular a period doubling bifurcation as described by \citet{MB90}. It is the first time such frequencies are reported in RR\,Lyrae stars.
\item The star KIC\,5559631 (V783\,Cyg) presented in this Letter has its 
modulation cycle (27.6\,d) covered by the Q1 data. For the other Blazhko stars the 
modulation periods are longer. Forthcoming data with a longer time base will allow for a multitude of analyses: besides the classical ways of analyzing Blazhko stars (Fourier techniques, phase dispersion minimization), we will investigate the relative contribution of phase and amplitude modulation in RR\,Lyrae stars \citep{SJ09,Ben09}.
\item In our forthcoming analyses, we will search for second and higher overtone 
(radial) modes in the time-series photometry, as well as potential nonradial 
modes. 
\item Several authors \citep{lacl04,kol06,jur09,cle97,gor09} have found that the 
modulation over the Blazhko cycle can be very unstable. The long time base of the {\it Kepler} data will allow us to study the stability of pulsation and modulation with unprecedented precision.
The {\it Kepler} data also have the required precision to investigate irregularities in the pulsation of non-modulated RR\,Lyrae stars.
\item We will also investigate features connected to shock wave phenomena, such as the change in slope at mid-rising light and  the amplitudes and durations of the ÒhumpÓ and ÒbumpÓ \citep[see][]{gil88} in both stable RRab stars and during Blazhko cycles.
\item Theoretically predicted strange mode pulsation \citep{bk01}, 
which has not been observed in RR\,Lyrae stars yet, may also be present in the data. 
\item Further possible applications of the 3.5-5 yr {\it Kepler} data are the measurement the 
secular variation of the pulsation period(s), 
directly linked with the evolutionary status of a star \citep[see, e.g.,][]{por08}, 
and the detection of substellar companions with
the timing method, as was done with another class of horizontal branch
pulsators \citep[e.g.,][]{sil07}. 
\end{itemize}

\vspace{2mm}

\acknowledgments
Funding for this Discovery mission is provided by NASA's Science Mission Directorate. The authors thank the entire {\it Kepler} team without whom these results would not be possible.
KK and EG are supported by the Austrian Research Fund (FWF) project T359 and P19962.  RSz, JB, ZK and JN acknowledge the financial support of KvVM-MUI grant No. K-36-08-00031K. DWK acknowledges support by the UK Science and Technology Facilities Council. 

\vspace{2mm}

{\it Facilities:} \facility{The {\it Kepler} Mission}

\clearpage


\begin{thebibliography}{}
\bibitem[Akerlof et al. (2000)]{akerlof00}Akerlof, C., Amrose, S., Balsano, R., et al., 2000, AJ 119, 1901
\bibitem[Alcock et al. (2003)]{alc03}Alcock, C., Allsman, R., Alves, D. R., et 
al., 2003, ApJ 598, 597
\bibitem[Benk\H o et al. (2009)]{Ben09}Benk\H o, J., Papar\'o, M., Szab\'o, R., et 
al., 2009, AIP Conf. Proc. 1170, 273
\bibitem[Blazhko (1907)]{Bla07} Blazhko, S.N., 1907, Astron. Nachr. 175, 325
\bibitem[Borkowski (1980)]{Bor80} Borkowski, K.J., 1980, Acta Astron. 30, 393
\bibitem[Buchler \& Koll\'ath (2001)]{bk01}Buchler, J.~R. \&  Koll\'ath, Z., (2001) ApJ 555, 961
\bibitem[Buchler \& Moskalik (1992)]{BM92}Buchler, J.~R. \& Moskalik, P., 1992, ApJ, 391, 736 (section 3.5)
\bibitem[Chadid et al. (1999)]{Cha99} Chadid, M., Kolenberg, K., Aerts, C., \& Gillet, D., 1999, A\&A 352, 201
\bibitem[Chadid et al. (2004)]{Cha04} Chadid, M., Wade, G. A., Shorlin, S. L. S., 
Landstreet, J. D., 2004, A\&A 413, 1087
\bibitem[Chadid et al. (2009)]{Cha09} Chadid, M., Benk\H o, J., Szab\'o, R., et al. 
2009, A\&A, accepted
\bibitem[Clement \& Goranskij (1999)]{cle97} Clement, C. \& Goranskij, V.P., 1999, ApJ 
513, 767
\bibitem[Csubry (2002)]{csu02}Csubry, Z., 2002, 2nd Workshop of Young Researchers in Astronomy
and Astrophysics, Ed. E. Forg\'acs-Dajka,
Publ. Astron. Dep. E\"otv\"os Lor\'and Univ. 12, 117
\bibitem[Dziembowski \& Mizerski (2004)]{DzM04}Dziembowski, W.A. \& Mizerski, T., 
2004, Acta Astron. 54, 363
\bibitem[Gillet \& Crowe (1988)]{gil88}Gillet, D. \& Crowe, R.A., 1988, A\&A 199, 242
\bibitem[Goranskij et al. (2009)]{gor09} Goranskij, V., Clement, C., \& Thompson, M., Proceedings of the Kukarkin Centenary Meeting, Eds. C. Sterken, in press
\bibitem[Gruberbauer et al. (2007)]{Gru07}Gruberbauer, M., Kolenberg, K., Rowe, 
J., et al. 2007, MNRAS 379, 1498
\bibitem[Hartman et al. (2004)]{hat04}Hartman, J.~D. et al., 2004, AJ 128, 1761
\bibitem[Jurcsik et al. (2006)]{jur06} Jurcsik, J., Szeidl, B., S\'odor, \'A., 
2006, AJ 132, 61
\bibitem[Jurcsik et al. (2009a)]{JS09} Jurcsik, J., S\'odor, \'A., Szeidl, B., et al., 2009a, MNRAS 393, 1553
\bibitem[Jurcsik et al. (2009b)]{jur09} Jurcsik, J., S\'odor, \'A., Szeidl, B., et al. 2009b, MNRAS 400, 1006
\bibitem[Kholopov et al. (1994)]{kho94}Kholopov P.N., Samus N.N., Frolov M.S, et 
al., 1994, General Catalogue of Variable Stars (GCVS)
\bibitem[Kolenberg (2002)]{Kol02}Kolenberg, K., PhD thesis, University of Leuven, Belgium
\bibitem[Kolenberg \& Bagnulo (2009)]{KB09}Kolenberg, K., \& Bagnulo, S., 2009, 
A\&A 498, 543
\bibitem[Kolenberg et al. (2006)]{kol06}Kolenberg, K., Smith, H. A., Gazeas, K. 
D., et al., 2006, A\&A 459, 577
\bibitem[Koll\'ath et al. (2002)]{Koll02}Koll\'ath, Z., Buchler J.R., Szab\'o, R., Csubry, Z., 2002, A\&A 385, 932
\bibitem[Kov\'acs (2009)]{Kov09} Kov\'acs, G., 2009, AIP Conf. Proc. 1170, 261
\bibitem[LaCluyz\'e (2004)]{lacl04}LaCluyz\'e, A., Smith, H.A., Gill, E.-M., et 
al., 2004, AJ 127, 1653
\bibitem[Lenz \& Breger (2005)]{LB05}Lenz, P. \& Breger, M., 2005, CoAst 146, 53
\bibitem[Mizerski (2003)]{Miz03}Mizerski, T., 2003, Acta Astron. 53, 307
\bibitem[Moskalik (1986)]{Mos86}Moskalik, P., 1986, Acta Astron. 36, 333
\bibitem[Moskalik \& Buchler (1990)]{MB90}Moskalik \& Buchler, 1990, ApJ 355, 590
\bibitem[Moskalik \& Buchler (1991)]{MB91}Moskalik \& Buchler, 1991, ApJ 366, 300
\bibitem[Moskalik \& Poretti (2003)]{MP03}Moskalik, P. \& Poretti, E., 2003, A\&A 
398, 213
\bibitem[Moskalik \& Olech (2008)]{MOl08} Moskalik, P., \& Olech, A., 2008, Comm. Asteroseis. 157, 345
\bibitem[Olech et al. (1999)]{Ol99}Olech, A., Ka\l{}u\.zny, J., Thompson, I.B., 1999, AJ 118, 442
\bibitem[Olech \& Moskalik (2009)]{OlM09}Olech, A., \& Moskalik, P., 2008, A\&A 494, L17
\bibitem[Pigulski et al. (2009)]{asas08} Pigulski, A. et al., 2009, Acta Astron. 59, 33
\bibitem[Pojma\'nski \& Maciejewski (2004)]{pm04}Pojma\'nski, G. \& Maciejewski, G. (2004) Acta Astron. 54, 153
\bibitem[Poretti et al. (2008)]{por08}Poretti, E., Le Borgne, J. F., Vandenbroere, J., et al., 2008,  Memorie della Societa Astronomica Italiana 79, 471
\bibitem[Reegen (2007)]{ree07}Reegen, P., 2007, A\&A 467, 1353
\bibitem[Roberts et al. (1987)]{Rob87}Roberts, D. H., Lehar, J., Dreher, J. W., 1987, AJ 93, 968
\bibitem[Samus et al. (2002)]{gcvs}Samus, N.~N., Goranskii, V.~P., Durlevich, O.~V., et al., 2002, Astr. Lett., 28, 174
\bibitem[Silvotti et al. (2007)]{sil07}Silvotti, R., Schuh, S., Janulis, R., et al., 2007, Nature 449, 189
\bibitem[Shibahashi (2000)]{Shi00}Shibahashi, H., 2000, ASP Conference Series, 
Vol. 203, 299
\bibitem[Stothers (2006)]{Sto06}Stothers, R.B., 2006, ApJ 653, 73,
\bibitem[Szab\'o et al. (2009)]{sza09}Szab\'o, R., Papar\'o, M., Benk\H o, J.M., et 
al., 2009, AIP Conf. Proc 1170, 291
\bibitem[Szeidl(1988)]{Sze88}Szeidl B. 1988, in Multimode Stellar Pulsations, 
Proc. Workshop Budapest 1987, 45
\bibitem[Szeidl \& Jurcsik (2009)]{SJ09} Szeidl, B., \& Jurcsik, J., CoAst, in 
press, {\it arXiv:0906.3987}
\bibitem[Stellingwerf (1987)]{Ste78}Stellingwerf, R.~F., 1978, ApJ 224, 953
\bibitem[Van Hoolst, Dziembowski \& Kawaler (1998)]{VH98}Van Hoolst, T., 
Dziembowski, W.A., Kawaler, S.D., 1998, MNRAS 297, 536

\end{thebibliography}
\end{document}